\documentclass{vgtc}                          

\ifpdf
  \pdfoutput=1\relax                   
  \usepackage{graphicx}                
  \DeclareGraphicsExtensions{.pdf,.png,.jpg,.jpeg} 
\else
  \usepackage{graphicx}                
  \DeclareGraphicsExtensions{.eps}     
\fi%

\graphicspath{{figures/}{pictures/}{images/}{./}} 

\usepackage{microtype}                 
\PassOptionsToPackage{warn}{textcomp}  
\usepackage{textcomp}                  
\usepackage{mathptmx}                  
\usepackage{times}                     
\usepackage{cite}                      
\usepackage{tabu}                      
\usepackage{booktabs}                  
\usepackage{color}

\onlineid{1699}

\vgtccategory{Research}

\vgtcinsertpkg

\title{P-Reverb: Perceptual Characterization of Early and Late Reflections for Auditory Displays}

\author{Atul Rungta\thanks{e-mail: rungta@cs.unc.edu}\\ %
       \parbox{1.4in}{\scriptsize \centering University of North Carolina \\Chapel Hill} %
\and Nicholas Rewkowski \thanks{e-mail: nr@unc.edu}\\ %
     \parbox{1.4in}{\scriptsize \centering University of North Carolina \\Chapel Hill}
\and Roberta Klatzky\thanks{e-mail: klatzky@cmu.edu}\\ %
     \parbox{1.4in}{\scriptsize \centering Carnegie Mellon University}
\and Dinesh Manocha\thanks{e-mail: dm@cs.umd.edu}\\ %
     \parbox{1.4in}{\scriptsize \centering University of Maryland \\ College Park}
\and \\URL: \url{ http://gamma.cs.unc.edu/preverb}}

\newcommand{\changes}[1]{\textcolor{black}{#1}}

\teaser{
  \centering
  \includegraphics[width=1\textwidth]{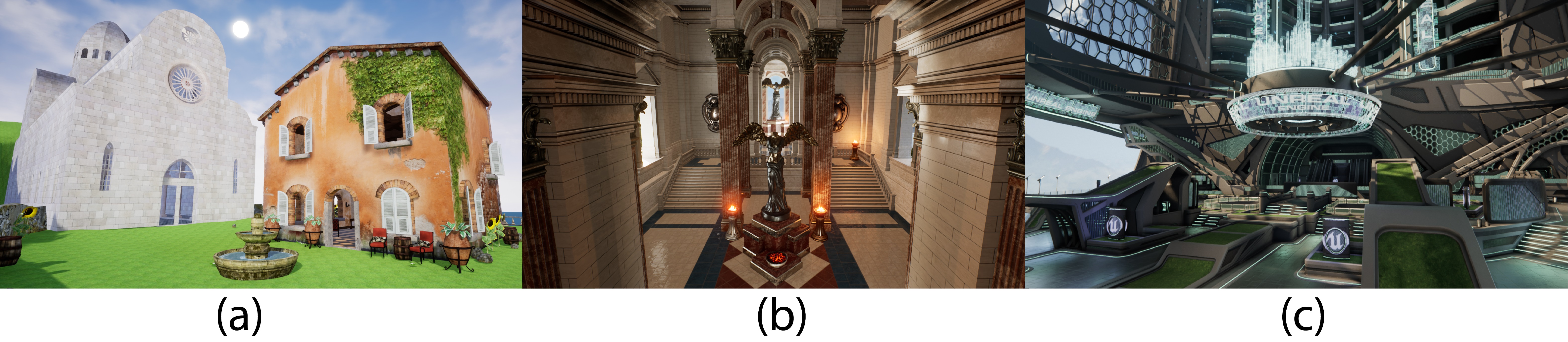}
  \caption{We highlight the performance of our $P-Reverb$ metric to generate reverberation effects in these complex scenes: (a) Tuscany, (b) Sun Temple, and (c) Shooter Game. We conduct extensive user-evaluations to establish our perceptual $P-Reverb$ metric that predicts the $RT_{60}$ based on the $JND$ of the mean-free path $\mu$ of an environment. We highlight the application of our metric by efficiently \changes{precomputing} the $RT_{60}$ values in the environment using $P-Reverb$ and use it to parameterize a reverberation filter at runtime for interactive sound propagation.}
	\label{fig:teaser}
}

\abstract{We introduce a novel, perceptually derived metric ($P-Reverb$) that relates the just-noticeable difference (JND) of the early sound field (also called early reflections) to the late sound field (known as late reflections or reverberation). Early and late reflections are crucial components of the sound field and provide multiple perceptual cues for auditory displays. We conduct two extensive user evaluations that relate the JNDs of early reflections and late reverberation in terms of the mean-free path of the environment and present a novel $P-Reverb$ metric. Our metric is used to estimate dynamic reverberation characteristics efficiently in terms of important parameters like reverberation time ($RT_{60}$). We show the numerical accuracy of our $P-Reverb$ metric in estimating $RT_{60}$. Finally, we use our metric to design an interactive sound propagation algorithm and demonstrate its effectiveness on various benchmarks.}

\begin{document}



\maketitle


\section{Introduction} 
Sound rendering uses auditory displays to communicate information to a user. \changes{Harnessing a user's sense of hearing enhances the user's experience and provides a natural and intuitive human-computer interface. Studies have shown a positive correlation  between the accuracy or fidelity of sound effects and the sense of presence or immersion in  virtual reality~\cite{larsson2002better,dubois2009engineering,rungta2016psychoacoustic}. Sound is also  an important cue for perceiving distance~\cite{zahorik2005auditory} and orientating oneself in an  environment~\cite{wilson2007swan}}. 

The sound emitted from a source and reaching the listener can be \changes{broken down into three components, described in more detail below: direct sound, early reflections, and late reflections or reverberation (Fig. \ref{fig:IR_image}). All three components of the sound field have perceptual relevance and have been extensively studied in psychoacoustics.  Direct sound gives us an estimate of the loudness and the distance to the sound source~\cite {zahorik2001loudness}.  Early reflections (ERs) arrive later than the direct sound, often in a range from $5$ to $80$ milliseconds.
Late reflections or reverberation (LRs) are generated when the sound signal undergoes a large number of reflections and then decays as it is absorbed by the objects in the scene.} 

Because of the importance of different components of sound fields, there has been considerable work on simulating these effects and incorporating them into auditory displays. 
Some of the commonly used methods 
approximate the sound field using artificial reverberation filters, which use 
reverberation time $(RT_{60})$ to tune parametric digital filters~\cite{valimaki2012fifty}. These filters tend to have low computational requirements and are widely
used for interactive auditory displays~\cite{kleiner1993auralization}. However, finding the right parameters for reverberation filters can be time-consuming and current methods do not provide sufficient fidelity. Geometric sound propagation methods work under the assumption that sound travels in straight lines and can be modeled using ray tracing~\cite{krokstad1968calculating}.
This allows resulting algorithms to model sound's interaction with the environment as it undergoes reflection and scattering.  Many techniques have been proposed to accelerate ray tracing, and current methods can generate early reflections (ERs) and late reflections (LRs) at  interactive rates  in dynamic scenes using high-order ray tracing (e.g., more than $100$ orders of reflections)~\cite{schissler2017interactive}. In practice, high-order ray tracing can be expensive and current interactive systems use multiple CPU cores on desktop workstations. The most accurate methods for sound rendering are based on wave-based acoustics, which  directly solve the acoustic wave equation using numerical methods.
However, their precomputation and storage overheads are very high and current methods are only practical for lower frequencies~\cite{mehra2012efficient,mehra2013wave,raghuvanshi2010precomputed}. 

Many applications, including games, virtual environments, and multi-modal interfaces require an interactive sound rendering capability, i.e., $20$fps or more. Furthermore, these systems are increasingly used on game consoles or mobile platforms where computational resources are limited. As a result, we need faster techniques to generate ERs and LRs in dynamic scenes for high-fidelity sound rendering. In particular, LR computation can be a major bottleneck.



{\bf Main Results:} We present a novel, perceptually derived metric called $P-Reverb$ that relates the ERs to the LRs in the scene. Our approach is based on the relationship between the mean-free path ($\mu$) and reverberation (Eq. \ref{eqn:mfp}), and we use early reflections to numerically estimate the mean-free path of the environment. We conduct two extensive user evaluations that establish the just-noticeable difference (JND) of sound rendered using early reflections and late reflections in terms of the mean-free path. We derive our perceptually-based $P-Reverb$ metric by expressing the JNDs of early and late reflections in terms of the mean-free path. Moreover, our metric is used to \changes{efficiently} estimate the late reverberation parameter ($RT_{60}$). 
We have evaluated the accuracy of our perceptual metrics in terms of computing the mean-free paths and reverberation time and comparing their performance with prior algorithms based on analytic or high-order ray tracing formulations. The mean-free path is within $3\%$ and reverberation time is within $4.6\%$, which are within the JND values specified by ISO 3382-1~\cite{iso20093382}.  
Overall, we observe significant benefits using our $P-Reverb$ metric for fast evaluation of mean-free path and reverberation parameters for sound rendering and auditory displays. We have used for sound propagation and rendering in complex indoor scenes.


The rest of the paper is organized as follows. We give a brief overview of prior work in sound propagation and psychoacoustics in Section 2. We present our user evaluations establishing the $P-Reverb$ metric in Section 3. We provide validation results in Section 4 and describe our how our metric can be used in an interactive sound propagation system in Section 5.

\section{Background \& Related Work} 
In this section, we give an overview of prior work in sound propagation, psychoacoustic characteristics, and related areas.

\subsection{Reverberation}
Reverberation forms the late sound field and is generated by successive reflections as they diminish in intensity. Reverberation is regarded as a critical component of the sound field. Many acoustic parameters such as the reverberation time (RT60) and clarity index (C50 and C80) are used to characterize reverberation~\cite{kuttruff2016room}. 

\begin{figure}[t]
\centering
\includegraphics[width=0.9\columnwidth]{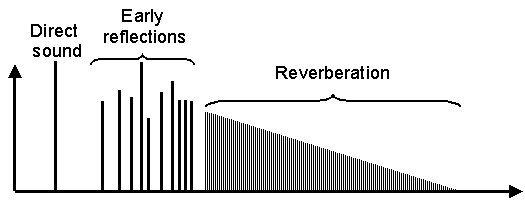}
\caption{We highlight the different components of the sound field. The sound directly reaching the listener is called the direct sound, the reflections that reach in the first $80$ ms are called early reflections (ERs), while the reflections following the early reflections that show a decaying exponential trend are called late reflections or reverberation (LRs). Our $P-Reverb$ metric presents a new perceptual relationship between ERs and LRs and we use it for fast sound rendering.}
 \vspace*{-0.1in}
\label{fig:IR_image}
\end{figure}

\subsubsection{Reverberation Time ($RT_{60}$):} $RT_{60}$ is defined as the time for the sound field to decay by $60$dB. A well-known expression used to compute the reverberation time is given by Sabine's formula, which gives the relationship between the RT60 of a room in terms of its volume, surface area, and the total absorption coefficients of the materials used: 
\begin{equation}
RT_{60} \approx 0.1611sm^{-1}\frac{V}{Sa},
\end{equation}
where V is the total volume of the room in $m^{3}$, S is total surface area in $m^{2}$, $a$ is the average absorption coefficient of the room surfaces, and $Sa$ is the total absorption in sabins. In this paper, we use $RT_{60}$ as the main reverberation parameter and use our $P-Reverb$ metric for fast computation in complex scenes.

\subsubsection{Mean-Free Path} The mean-free path (MFP) of a point in the environment is defined as the average distance a sound ray travels in  between collisions with the environment and is directly related to the $RT_{60}$ \cite{kuttruff2016room}:
\begin{equation}
RT_{60} = k\frac{\mu}{log(1 - \alpha)},
\end{equation}
where $k$ is the constant of proportionality, $\mu$ is the mean-free path, and $\alpha$ is the average surface absorption coefficient. A closed form expression ~\cite{bate1947mean} for computing the mean-free path is given by:
\begin{equation}
\mu = \frac{4V}{S},
\label{eqn:mfp}
\end{equation}
where $V$ is the volume of the environment and $S$ is the surface area. 
The mean-free path can be computed to a reasonable degree of accuracy by only considering the specular reflection paths in the scene \cite{vorlander2000room}. We use our $P-Reverb$ metric for fast computation of $\mu$ using only ER in complex scenes.

\subsection{Sound Propagation and Acoustic Modeling}
Artificial reverberators provide a simple mechanism to add reverberation to ``dry" audio, which has led to their widespread adoption in the music industry, virtual acoustics, computer games, and user interfaces. One widely used artificial reverberator was introduced by Schroeder~\cite{schroeder1961colorless} and it uses digital nested all-pass filters in combination with a parallel bank of comb filters to produce a series of decaying echoes. These filters require parameters such as reverberation time $(RT_{60})$ to tune the all-pass and comb filters. Geometric methods work on the underlying assumption of the rectilinear propagation of sound and use ray tracing to model the acoustics of the environment~\cite{krokstad1968calculating}. Other geometric methods include beam tracing~\cite{funkhouser1998beam} and frustum tracing~\cite{chandak2008ad}. In practice, ray tracing remains the most popular because of its relative simplicity and generality and because it can be accelerated on current multi-core processors. Over the years, research in ray tracing-based sound propagation has led to efficient methods to compute specular and diffuse reflections~\cite{schissler2014high} for a large number of sound sources~\cite{schissler2017interactive}.

\subsection{Early \& Late Reflections: Psychoacoustics}
Early reflections (ERs) have been shown to have a positive correlation with the perception of auditory spaciousness and are very important in concert halls. \cite{barron1971subjective,blauert1986auditory} showed that adding early reflections generated the effect of ``spatial impression'' in subjects. Early reflections are also known to improve speech clarity in rooms. \cite{bradley2003importance} showed that adding early reflections increased the signal-to-noise ratio and speech intelligibility scores for both impaired and non-impaired listeners. 
\cite{hartmann1983localization} showed that early reflections that come from the same direction as the direct sound reinforce localization, while those coming from the lateral directions tend to de-localize  the sources. 

Late reflections or reverberation (LRs) provide many perceptual cues. Source localization ability deteriorates in reverberant conditions, with localization accuracy decreasing in a reflecting room compared to the same absorbing room~ \cite{hartmann1983localization}. Reverberation has a negative impact on speech clarity and ~\cite{knudsen1932architectural} showed the reduction in the number of sounds heard correctly in the presence of reverberation.
Although reverberation decreases localization accuracy and speech intelligibility, it is known to have positive effects with respect to the perceived distance to a sound source in the absence of vision~\cite{zahorik2005auditory}. 

While there is considerable work on  separately characterizing the perceptual effects of ERs or LRs, we are not aware of any work that establishes any relationship between ERs and LRs. $P-Reverb$ is a metric that establishes the relationship between the respective JNDs and uses them for interactive sound rendering.

\subsection{Estimating Reverberation Parameters} Given the importance of reverberation to the overall sound field, multiple methods have been established to measure the reverberation parameters over the years. $(RT_{60})$, in particular, is considered to be the most important parameter in estimating reverberation and has been referred as the `The mother of all room acoustic parameters' \cite{skaalevik2010reverberation}. The most commonly used method to estimate reverberation time was given by Schroeder, and uses a backward time integration approach. \cite{ratnam2003blind} presents a method for blind estimation of $RT_{60}$ that does not require previous knowledge of sound sources or room geometry by modeling reverberation as an exponentially damped Gaussian white noise process. \cite{lollmann2008estimation} describes a method to estimate reverberation time using maximum likelihood estimator from noisy observations. \cite{vorlander1994comparison} presents a comparison of different methods for estimating $RT_{60}$.

\section{Perceptual Evaluations and P-Reverb} In this section, we describe two user evaluations that establish the just-noticeable difference (JND) for early and late reflections in terms of the mean-free path. Further, we show the relationship between the two JND values, thereby establishing our $P-Reverb$ metric. 

\subsection{Experiment I - Just-noticeable difference of ERs}
In this experiment, we seek to establish the just-noticeable difference $(JND_{er})$ of sound rendered using only direct and early reflections. In Experiment II, we establish the relationship between $JND_{er}$ and sound rendered using the full simulation (direct + early + late reverberation) $JND_{lr}$. 

\paragraph{Participants:} 106 participants took part in this web-based, online study. The subjects were recruited using a crowd-sourcing service. All subjects were either native English speakers or had professional proficiency in the language.

\paragraph{Apparatus:} The online survey was set up in Qualtrics. The impulse responses were generated using an in-house, realtime, geometric sound propagation engine written in C++, while the convolutions to generate the final sounds were computed using MATLAB.

\paragraph{Stimuli:} \changes{The stimuli were sound clips derived from $7$ cube-shaped rooms with increasing edge lengths such that their MFPs (Eq. \ref{eqn:mfp}) varied from $2 - 2.2$m in increments of $0.033$m. The range of lengths was chosen with the experimental goal in mind, namely, to extract a psychophysical function showing a gradient in perceived sound similarity relative to edge-length difference.  The walls of the rooms had reflectivity similar to that of an everyday room. The source was a sound of clapping, which was chosen because it represents a broadband signal.} The clips were filtered in 4 logarithmically spaced frequency bands ($0 - 176$Hz, $176 - 775$Hz, $775 - 3408$Hz, and $3408 - 22050$Hz) to evaluate the effects of frequency on $JND_{er}$. Each of these $4$ filtered clapping sounds was convolved with the early impulse responses of the $7$ rooms. The final sound clip was around $4$ seconds long and contained $3$ distinct parts: $1.5$ seconds of the clapping sound in Room 1 ($\mu = 2m$), $1$ second of silence, and another $1.5$ seconds of clapping in a second room drawn from $1 - 7$ ($\mu = [2 to 2.2]m$). \changes{All sounds were recorded assuming that the listener and the source were located at the origin $(0,0,0)$. Given this symmetry, the sounds were rendered in mono with both speakers playing the same sound.} 

\paragraph{Design \& Procedure:} \changes{To estimate the JND, our experiment used the method of constant stimuli~\cite{gescheider2013psychophysics} with a within-subject design. A stimulus comprised a sound clip containing Room 1 and one of the 7 possible comparison rooms (including Room 1). For each clip the subjects heard, they were asked to identify if the first clapping sound seemed to be different from the second clapping sound by selecting yes or no. Note this is a similarity judgment, not a discrimination.  A block of judgments consisted of $28$ clips ($4$ frequencies x $7$ comparison rooms paired with Room 1). A block was repeated $5$ times, giving a total of $140$ clips (4 frequencies x 7 possible rooms paired with Room 1 x 5 blocks). The ordering of the clips was randomized within a block. Each subject judged all $140$ stimuli.  Before starting the experiments, subjects listened to a sample clip for familiarization. The subjects were required to have a pair of ear-buds/headphones to take part in the study, which took an average of $25$ minutes to complete.}

\begin{figure}[h]
	\includegraphics[width=0.99\columnwidth]{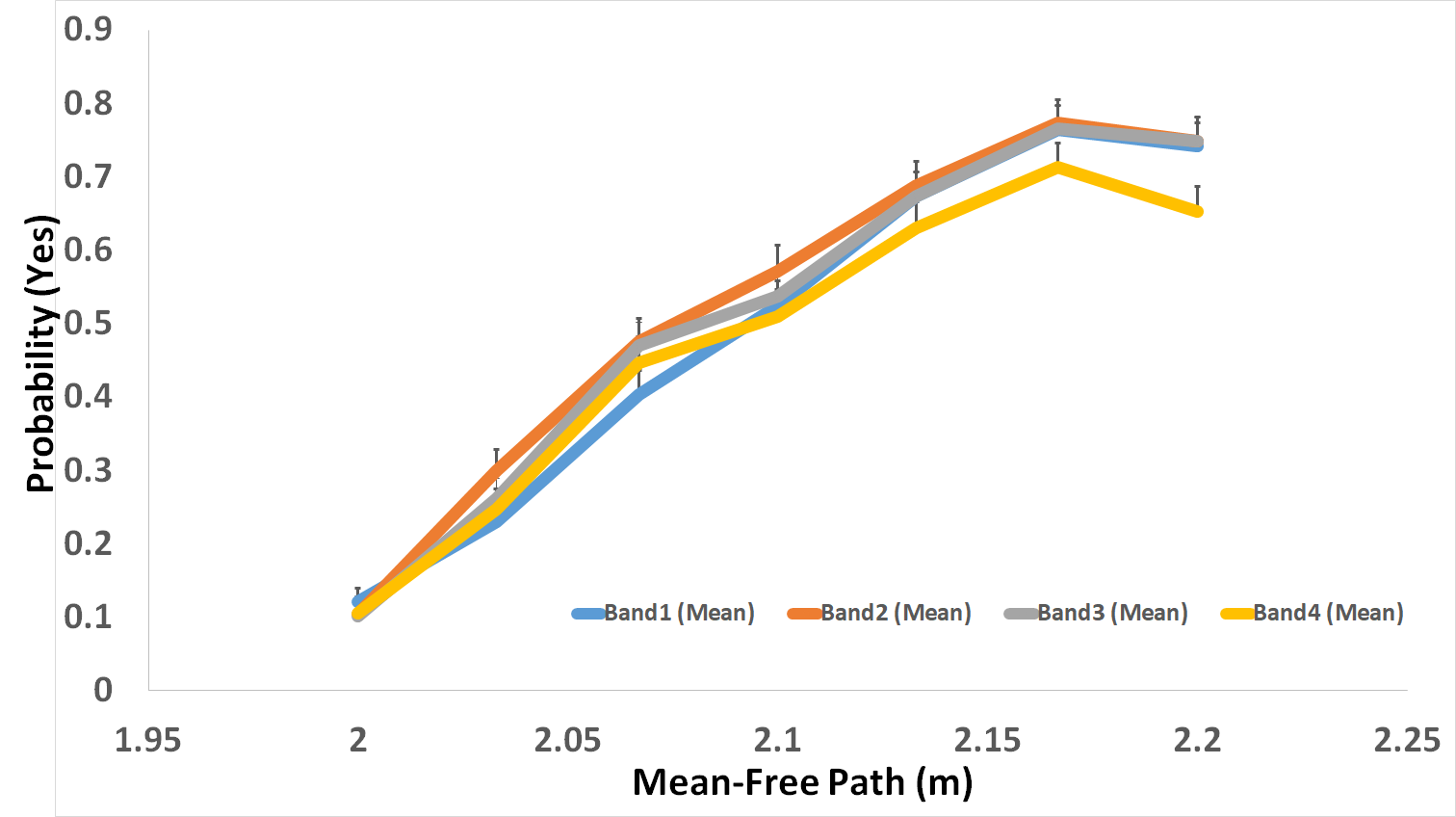}
\caption{The psychometric function for sound rendered using the early reflections for the 4 frequency bands. The Y-axis shows the proportion of responses indicating the sounds were different. \changes{We see a clear, linear trend between increasing $\mu$ and the probability of responding different, until the last room $\mu = 2.2$, where the responses seem to flatten out. }}
\label{fig:MFP_JND}
\end{figure}

\paragraph{Results \& Analysis:}  Fig. \ref{fig:MFP_JND} shows the proportion of responses in which rooms were judged as sounding different, over all participants, as a function of the comparison level of $\mu$. The first data point corresponds to rooms that were objectively identical, providing a baseline.  The data essentially increases linearly with a larger $\mu$, showing greater discrimination up to Room $6$ ($\mu = 2.17$), after which ($\mu = 2.2$) the discriminatory ability seems to taper off. The standard errors are low and consistent, indicating the robustness of the results. 

\begin{figure}[t]
	\includegraphics[width=0.95\columnwidth]{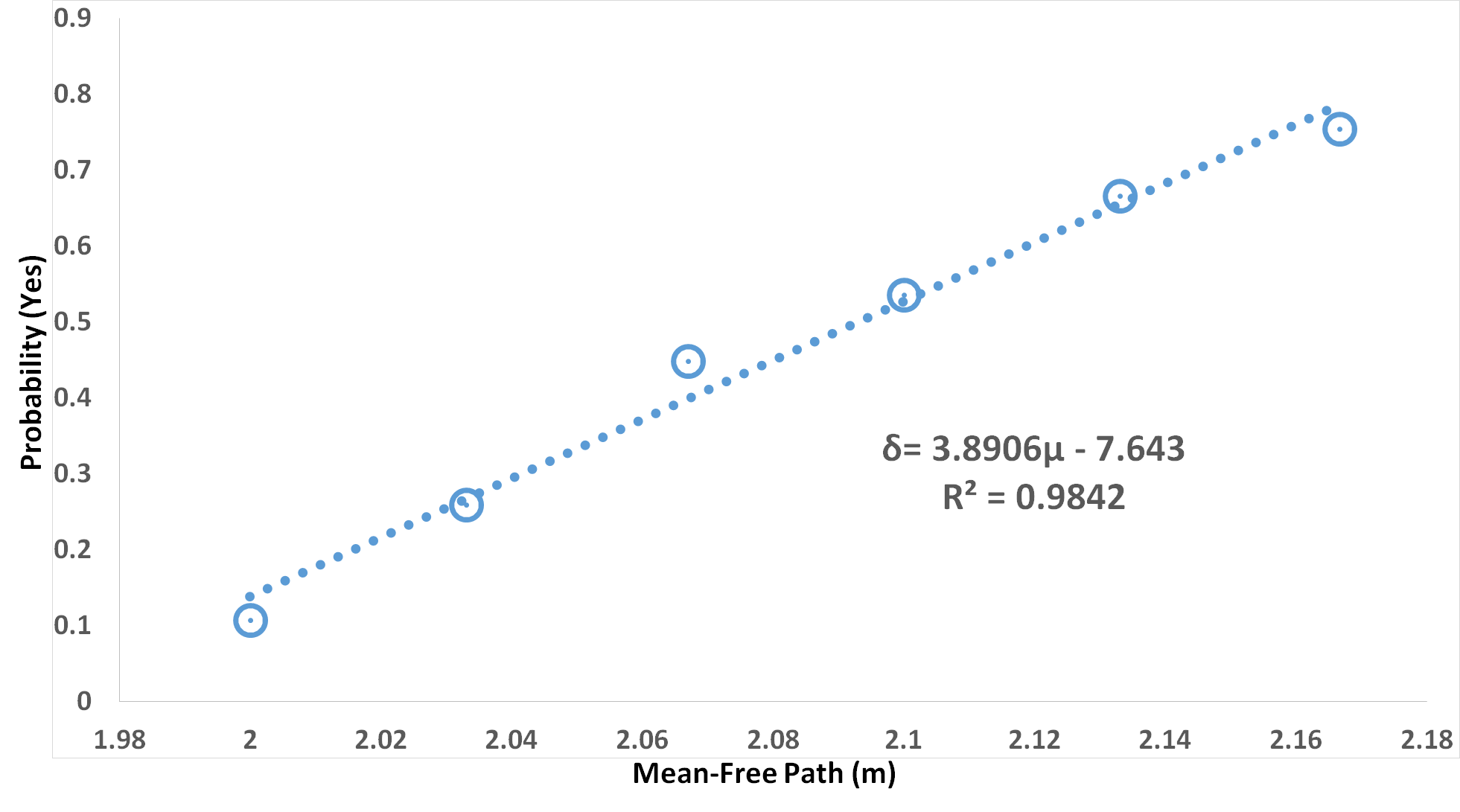}
\caption{The average JND over the frequency bands. The Y-axis shows the proportion of responses indicating that a difference was \changes{judged. The psychophysical function is essentially linear, showing that the probability of judging the sounds as different increases linearly with the increasing mean-free paths of the rooms. }}
\label{fig:MFP_JND_AVG}
\end{figure}

An interesting observation is the near-invariance of subjects' ability to discriminate across the frequency bands. This was verified by an ANOVA analysis with factors of edge length (or $\mu$) and frequency. The analysis showed significant main effects for both factors of edge length $F(6, 180) = 61.78, p < 0.05, \eta^{2}_{p} = 0.673 $ and frequency $F(3, 90) = 2.95, p = 0.037, \eta^{2}_{p} = 0.09 $. The interaction between edge length and frequency was also significant $F(18, 540) = 1.66, p = 0.04, \eta^{2}_{p} = 0.052 $, reflecting that the performance decrement at the largest edge length is slightly greater for frequency band $4$.  However, the $\eta^{2}_{p}$ values are very low for effects involving frequency. Thus, while the effects of frequency show statistical significance, they are small in effect size and do not reflect consistent variation in frequency across edge length (or $\mu$).  Therefore, for the purposes of constructing an overall rule, using data averaged over the  frequencies is  a valid simplification, particularly if the largest value of edge length is excluded. Fig. \ref{fig:MFP_JND_AVG} shows the results averaged over the frequency for Rooms $1 - 6$. As shown in the figure, the data fits a linear function well, with $R^{2} = 0.98$.  
Given our linear fit:
\begin{equation}
\delta = 3.89\mu - 7.5
\label{eq:linear_jnd_fit},
\end{equation}
we can easily estimate the $JND_{er}$ by considering the MFP values ($\mu_{JND}$) where the subjects successfully discriminated the sounds 50\% of the time given by $\mu_{JND} = \mu_{50\%} - \mu_{room1} = 2.06 - 2 = 0.06 m$. This tells us that a change in $\mu$ greater than 0.06m would result in perceptually differentiable sounds when using \textit{early impulse responses}, but it doesn't necessarily indicate if the relationship holds if the sounds were rendered using the \textit{full impulse response} (LR). This led us to conduct the next experiment to establish the relationship between the JND of early reflections $(JND_{er})$ and the JND of full impulse response or late reverberation $(JND_{lr})$.

\subsection{Experiment II - Relationship between $JND_{er}$ \& $JND_{lr}$} 
Once we have established the perceptibility threshold or $JND_{er}$ of ERs, we need to relate this to the $JND_{lr}$ of LRs. Our goal is to use these relationships to cluster points $p \in P$ with similar reverberation characteristics. We conducted another user study based on the results of the first study, described above. 

\paragraph{Participants} $31$ participants took part in this online, web-based study. The subjects were recruited using the same crowd-sourcing service as in the previous experiment. All subjects were either native English speakers or had professional proficiency in the language.

\paragraph{Apparatus} The apparatus was the same as in Experiment I. The full impulse responses were generated using our in-house, realtime, geometric sound propagation engine written in C++, with the convolutions being computed using MATLAB.

\paragraph{Stimuli} The sound source used was the same as in the previous experiment, filtered for the same logarithmically-spaced frequency bands. Given our goal of establishing the relationship between $JND_{er}$ and $JND_{lr}$, we use our previously computed psychometric  function (Eq. \ref{eq:linear_jnd_fit}), to compute the \textit{$6 \mu$ values corresponding to detection rates ranging from $0.2$ to $0.7$}. This gives us $6$ $\mu$ values that can then be used to compute the cube rooms' edge lengths using Eq. \ref{eqn:mfp}. These $6$ rooms and Room $1$ from the previous experiment serve as the environments in which the full impulse responses are computed. The material properties of the rooms were the same as in the previous experiment. Each sound clip in this case was about $6$ seconds because of the increased length of full impulse response, with $2.5$ seconds of clapping in Room $1$, followed by a second of silence, followed by $2.5$ seconds of clapping in Rooms $2 - 7$. The total number of sound clips was $14$0, as before ($4$ frequency bands x $7$ rooms x $5$ blocks). The ordering of the sound clips was randomized within each block.

\paragraph{Design \& Procedure} The study design was the same as in the ER study. Before starting the study, the subjects were asked to listen to a sample sound clip from the $28$ clips computed above for familiarization. The source and listener locations in the rooms were located at $(0,0,0)$. The sound was rendered in mono. The subjects took an average of $30$ minutes to complete the study.

\begin{figure}[h]
	\includegraphics[width=0.95\columnwidth]{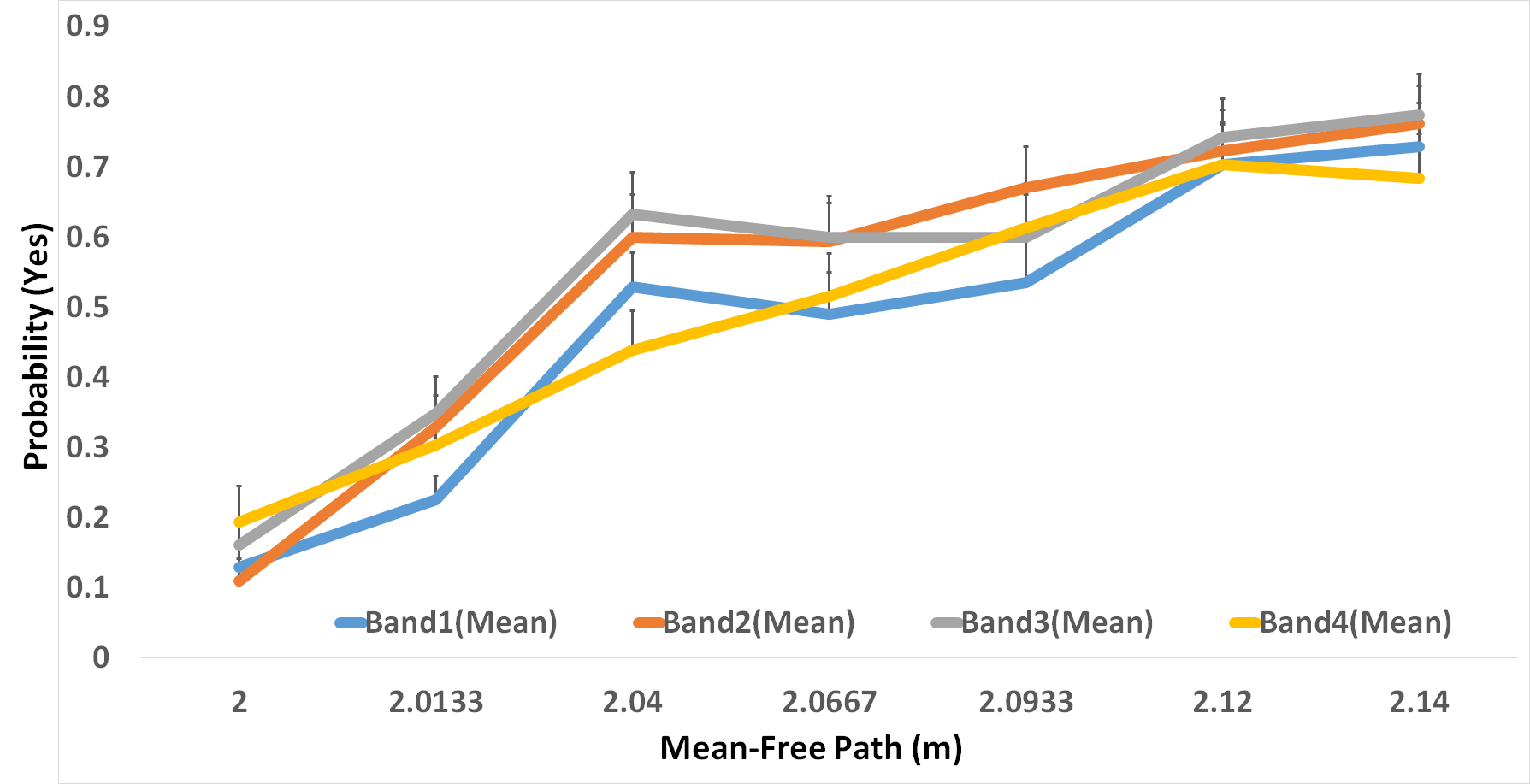}
\caption{The psychometric function for sound rendered using the full impulse response $(LR)$ for the $4$ frequency bands. The Y-axis shows the proportion of responses indicating sounds were judged to be different. In this case, we observe more variability for the different frequency bands, which could be attributed to the greater sensitivity of human hearing to a more accurate signal (compared to the less accurate ER signal). Overall, however, the responses can be modeled as a linear function with reasonable accuracy.}
\label{fig:JND_RT_60}
\vspace*{-0.15in}
\end{figure}

\begin{figure}[h]
	\includegraphics[width=0.95\columnwidth]{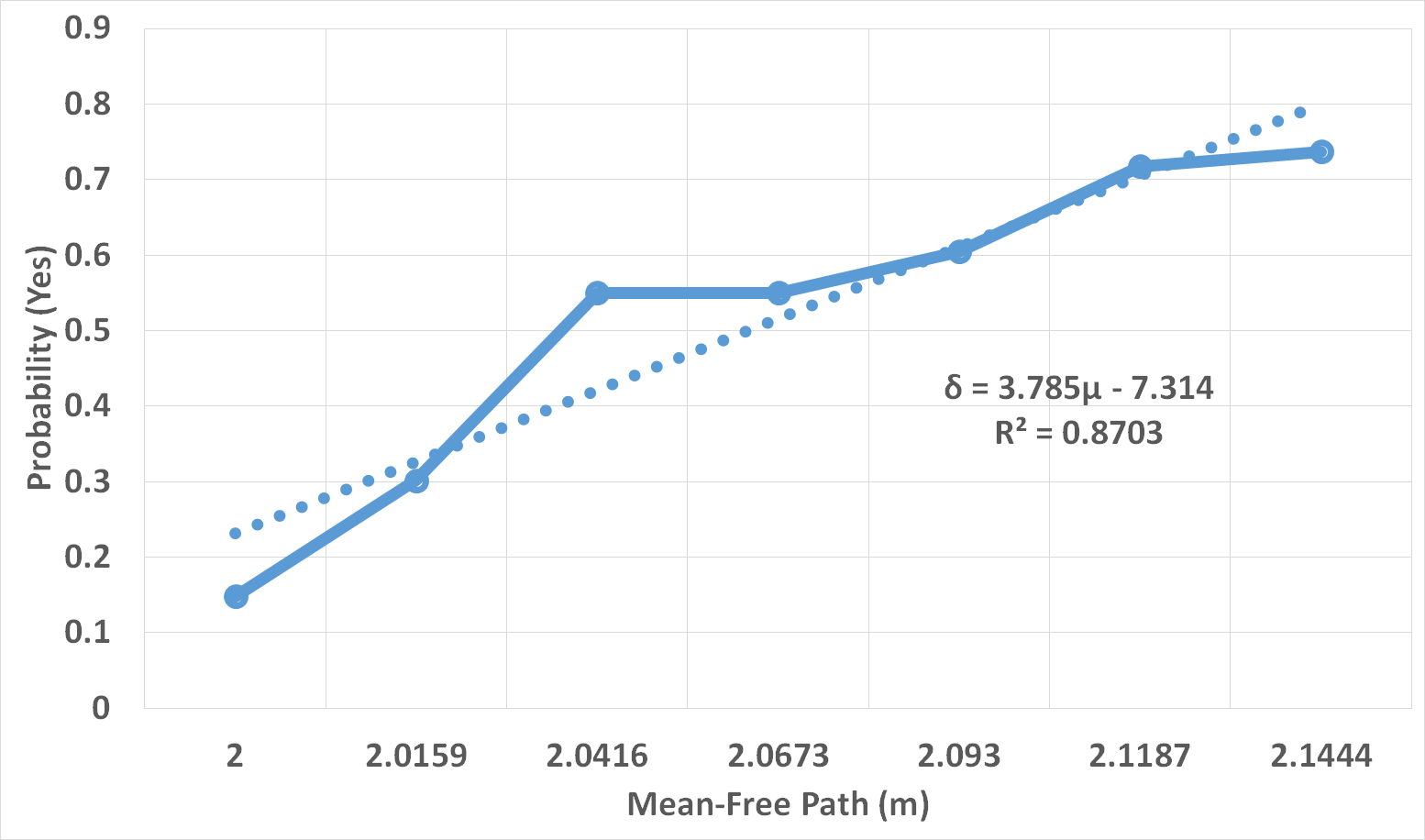}

\caption{The average JND over the frequency bands for the full impulse response signal. \changes{The Y-axis shows the proportion of responses indicating a judgment of difference}. The psychophysical function is not as linear as the early reflection signal, but a linear function approximates the subject responses reasonably well ($R^2 = 0.87$), accounting for most of the variability.}
\label{fig:JND_RT60_AVG}
\end{figure}

\paragraph{Results \& Analysis} Fig. \ref{fig:JND_RT_60} \changes{shows the proportion of responses judging the sounds as different, as a function of increasing $\mu$ or edge-length}. As before, we performed an ANOVA to assess the effect of edge length and frequency. The analysis showed significant main effects for edge length $F(6, 180) = 61.78, p < 0.05, \eta^{2}_{p} = 0.673 $ and frequency $F(3, 90) = 2.95, p = 0.037, \eta^{2}_{p} = 0.09 $. The interaction between edge length and frequency was also significant $F(18, 540) = 1.66, p = 0.04, \eta^{2}_{p} = 0.052 $. Again, the effect size for terms involving frequency was low, allowing us to average the responses for the frequency bands. Fig. \ref{fig:JND_RT60_AVG} shows the values averaged for the $4$ frequency bands.

\subsection{$P-Reverb$ Metric} 
\begin{figure}[h]
	\includegraphics[width=0.95\columnwidth]{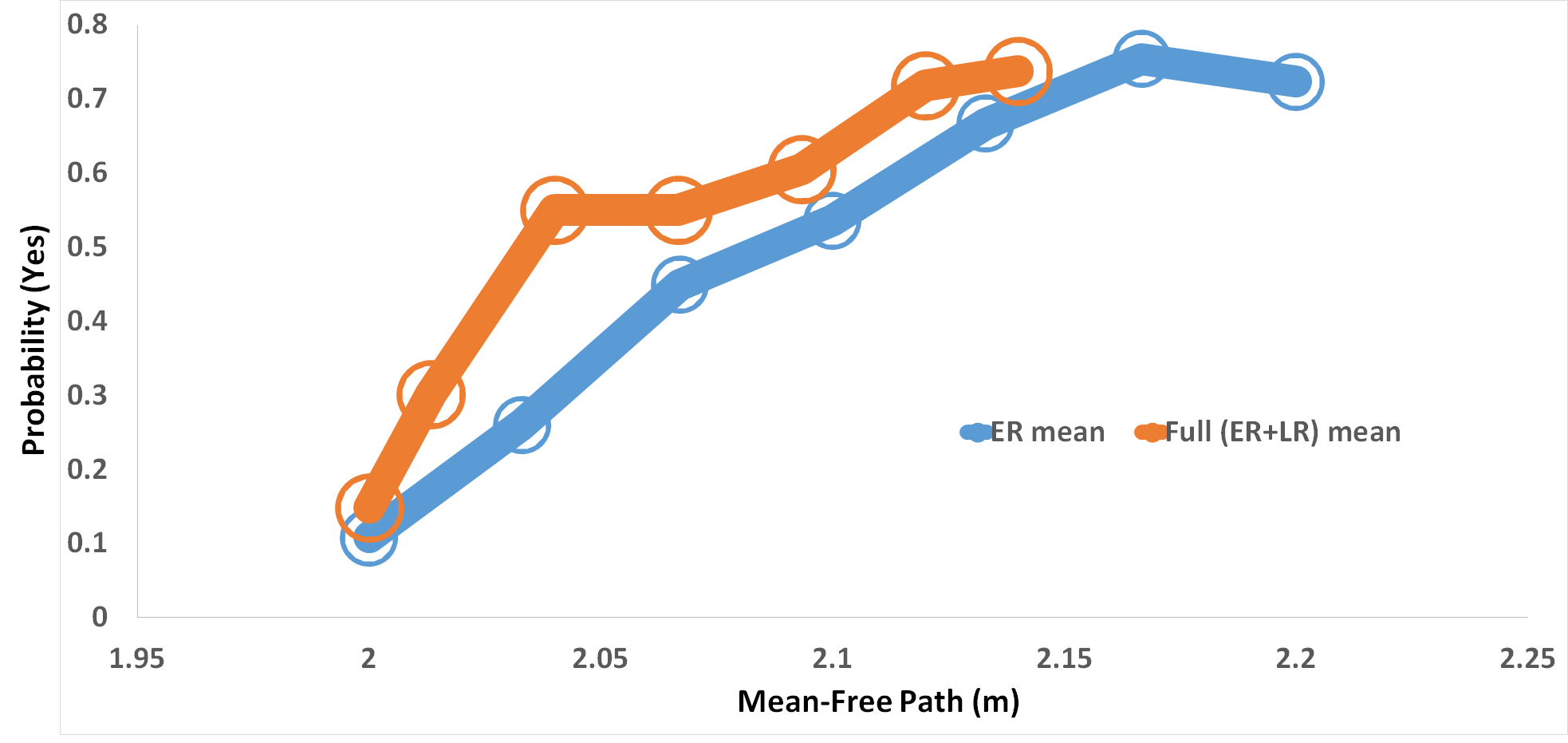}

\caption{This plot shows the overlaid psychometric functions for signals
rendered using the early reflections (blue) and full impulse response (orange). Note that the first data point corresponds to differences being reported when the stimuli are objectively identical.  Although the full impulse data shows a greater departure from a linear relationship beyond that point, the results are similar to the early reflection function, offset by a constant, allowing us to establish a simple, linear relationship between $JND_{er}$ and $JND_{lr}$ in terms of the mean-free path.}
\label{fig:RT60_VS_MFP}
\end{figure}

Fig. \ref{fig:RT60_VS_MFP} shows the relationship between the sounds rendered using only the early responses and the sounds rendered using the full impulse response. Note that the first point for both functions corresponds to two identical stimuli, and no difference is expected.  However, beginning at the smallest edge lengths where objectively different stimuli were presented, the figure shows that the subjects were more likely to \changes{differentiate} between sounds rendered with the full impulse response than they were with sounds rendered using only the early reflections. A difference in \changes{difference judgments is expected,} because the  full impulse response conveys more information about the space and is supposed to enable better perceptual differentiation than the early impulse response, thus giving a lower JND for the full impulse response, i.e., $JND_{lr} < JND_{er}$ .

\begin{figure}[h]
	\includegraphics[width=1\columnwidth]{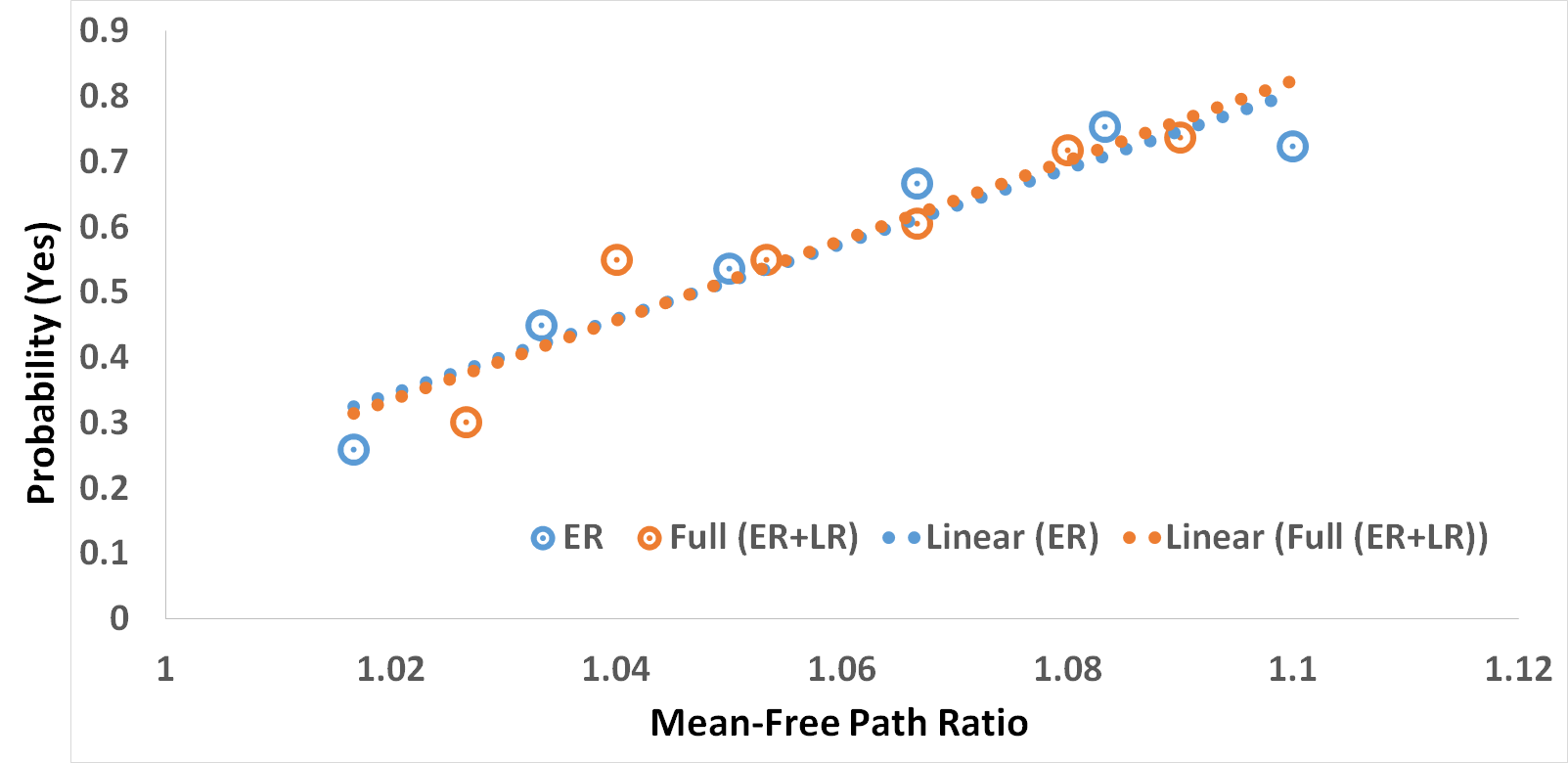}

\caption{The psychometric function with a constant offset adjustment. We consider the ratio of the mean-free path for the different rooms to the mean-free path of Room 1. The resulting linear fits for the two cases (early reflections and full impulse response) coincide once a constant offset of 0.02 is added to the ratio for the full impulse response. This highlights the accuracy of our model.}
\label{fig:MFP_ADJ}
\end{figure}

To establish a mathematical relationship between the two JNDs, we consider the ratio of the mean-free paths in both cases. The resulting figure is shown in Fig. \ref{fig:MFP_ADJ}. The linear fits are almost coincident after adding a constant offset of $0.02$, i.e. 
\begin{equation}
\frac{\mu_{JND_{lr}}}{\mu_{1}} + 0.02 = \frac{\mu_{JND_{er}}}{\mu_{1}},
\end{equation}
which gives a simple relationship between the two JND values:

\begin{equation}
\mu_{JND_{lr}} = \mu_{JND_{er}} - 0.02 \mu_{1},
\label{eq:jnd_rel}
\end{equation}
where $\mu_{1}$ is the mean-free path in Room $1$ = $2$m. Hence $\mu_{JND_{lr}} = \mu_{JND_{er}}- 0.04$ is the simple mathematical relationship or $P-Reverb$ for the JND values of the two signals. Given $\mu_{JND_{er}} = 0.06 m$ as derived above, we can easily compute the value of $\mu_{JND_{lr}}$ as being $0.02m$ for a reference room (Room 1) $\mu = 2m$, \textit{giving us the percentage change ($\frac{\mu_{JND_{er}}}{\mu_{Room 1}} = 1 \%$) in the mean-free path values that constitute the JND for late reverberation}, when using early reflections.

It turns out that  Eq. \ref{eq:jnd_rel} can be interpreted as a ``first-order" approximation to a function that expresses the mathematical relationship between two multi-dimensional perceptual phenomena that are dependent on frequency, edge length, method of rendering, material parameters, etc. However, any function that accounted for the small frequency dependencies in the observed psychometric data and accommodated the effects of more complex environments and material parameters would have to be substantially more complicated than the linear relationship that we derive here. The value of the present formulation lies in its reasonable approximation of the observed effects with only one derived parameter. 

We would also like to note that, although psychometric functions are usually fitted using sigmoid functions, our design did not require us to do so. A sigmoid function approach would have been suitable had we started with a value somewhere in the middle and taken a range of values above and below. This would have yielded two end-cases with the non-standard stimulus being judged smaller 100 \% of the time; similarly, the larger non-standard stimulus would be judged as such 100 \% of the time. In our approach, however, we never tested anything smaller than the standard, which led us to values that rose to the ceiling. Consequently, a linear fit to this function accounted for most of the variance $(93 \%)$. A better fit could be achieved using a quadratic fit (accounting for $99 \%$ of the variance), but at the expense of adding a parameter. A sigmoid function, too, would add another parameter without yielding much gain. Therefore given the fact that our linear fit accounts for most of the variability, we chose to not use a sigmoid fit. 


\section{Results \& Evaluation}
Our approach consists of two primary numerical steps: computing the mean-free path $(\mu)$ using early reflections (ERs), and predicting $RT_{60}$ using our perceptually established $P-Reverb$ metric. We first validate the use of early reflections (ERs) to compute the mean-free path ($\mu$) in various environments. Next, we highlight the validation of the $P-Reverb$ metric in terms of its accuracy in predicting $RT_{60}$. 



\begin{figure*}[h]
    \center
	\includegraphics[width=1\textwidth]{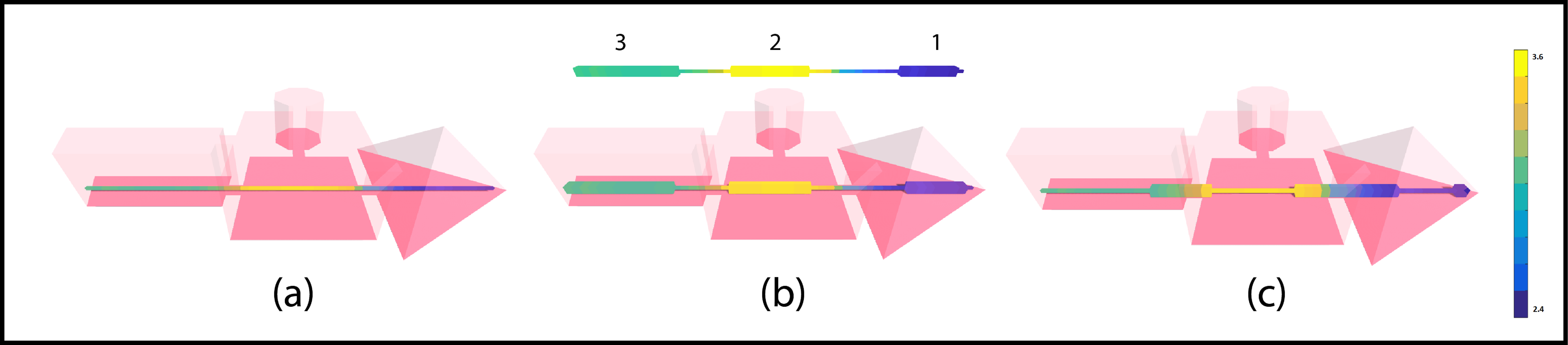}
\caption{ We highlight the application of $P-Reverb$ metric to predict variations in $RT_{60}$ in a scene composed of interconnected rooms of different shapes and volumes: (a) shows the variation in $\mu$ along a path that goes through three rooms with volumes 135 $m^3$, 256$m^3$, and 125 $m^3$ from left to right; (b) shows three regions $(1,2,3)$ along the path roughly corresponding to the three rooms, where $\mu$ changes within the JND specified by the $P-Reverb$ metric. This indicates that the reverberation in these regions would vary imperceptibly, as is indicated by the uniformity of the $\mu$ values; (c) shows rapidly varying $\mu$ values as one approaches the apertures between the connected rooms, indicating that $RT_{60}$ would also vary rapidly. This is expected because the geometry varies rapidly in these regions and validates the accuracy of our perceptual metric $P-Reverb$.}

\label{fig:preverbvalidation}
\end{figure*}

\subsection{Mean-Free Path Computation} Our $P-Reverb$ metric depends on the numerically computed mean-free values that are computed using early reflections. The mean-free path is the average distance a sound ray would travel between collisions and we use ERs to estimate this distance. As mentioned, Eq. (\ref{eqn:mfp}) can be used to compute mean-free path values in terms of the volume ($V$) and surface area ($S$).  Table \ref{table:MFP_Validation} highlights the accuracy of our computed mean-free path values ($\mu_{er}$) as compared to the analytical values given by Eq. \ref{eqn:mfp}. We use $500$ rays and 20 bounces for each ray to compute our $\mu_{er}$ value as:
\begin{equation}
\mu_{er} = \frac{\sum d_{i}}{n \times b},
\end{equation}
where $d_{i}$ is the distance traveled by a sound ray on the $i^{th}$ bounce, $n$ is the total number of rays, and $b$ is number of bounces per ray.

{\small
\begin{table}[h]
        \begin{tabular}{|r|c|c|c|c|c}
        \hline
      {$\mathbf{Shape}$} & {$\mathbf{Dim. (m)}$} & {$\mathbf{\mu_{er} (m)} $} & {$\mathbf{\mu_{an} (m)}$} & {$\mathbf{\% error}$}  \\
        \hline
        
        Cube              &   5                       & 3.3    & 3.33 & 1\\      
        Rect. Prism       &   (2,3,4)                 & 1.87   & 1.85 & 1.3\\
        Sq. Pyramid       &   (2.8, 3) (b, h)         & 1.16   & 1.18 & 1.7\\
        Room with Pillars &   (5,6,12)                & 3.14   & 3.04 & 3\\
        \hline
        \end{tabular}
        \vspace*{0.1in}
\caption{\textbf{Mean-free path Computation:} We show the accuracy of computing $\mu_{er}$ using early reflections for differently shaped rooms. The closed-form expression in Eq \ref{eqn:mfp} gives us the analytical value for the mean-free paths in each of the rooms $\mu_{an}$. We observe that ERs can closely approximate the analytically obtained $\mu_{an}$. The Room with Pillars  is shown in Fig. \ref{fig:roomwithpillars}. Even for a scene with multiple obstacles, our method computes the mean-free path while inducing a maximum error of only $3\%$. }
\label{table:MFP_Validation}
\end{table}
}

\subsection{$RT_{60}$ using $P-Reverb$ Computation} The $P-Reverb$ metric predicts regions in a scene where the late reverberation is likely to vary imperceptibly. Conversely, it can estimate regions where the late reverberation would vary in a perceptually noticeable manner. We demonstrate the effectiveness of the $P-Reverb$ metric in finding regions of similar reverberation characteristics by considering a scene shown in Fig. \ref{fig:preverbvalidation}. The scene is composed of different interconnected rooms of varying shapes and volumes. Since reverberation is a function of the volume and shape of the room, it is likely to vary as one moves from one room to another. We consider a path that traverses three different connected rooms and compute the mean-free path along the path using  ERs. Fig. \ref{fig:preverbvalidation}(a) shows the variation in $\mu$ as we move along the path.  We group the regions along the path where $\mu$ varies within the JND threshold computed using our $P-Reverb$ metric (as shown in Fig. \ref{fig:preverbvalidation}(b)), as Regions $1$, $2$, and $3$. Based on the $P-Reverb$ metric, each such region is likely to have an imperceptible sound in terms of $RT_{60}$. We illustrate this in Table \ref{table:preverbvalidationtable}. The $\mu_{mean}$ corresponds to the average mean-free path value for the entire region and $Diff._{max}^{\mu}$ corresponds to the maximum difference from the $\mu_{mean}$ for all the points in that region (i.e., a measure of variance). The $RT_{60}^{mean}$ represents the average value of the reverberation times for the region, while $Diff._{max}^{RT_{60}}$ corresponds to the maximum difference from $RT_{60}^{mean}$. For regions where $\mu$ varies within the JND specified by the $P-Reverb$ metric, the $RT_{60}$ values vary within 5\% of the $RT_{60}^{mean}$. This is within established JND values for $RT_{60}$, as specified in ISO 3382-1 \cite{iso20093382} and correspond to imperceptible changes in late reverberation.

Fig. \ref{fig:preverbvalidation}(c) shows rapidly varying $\mu$ values, as one moves from one room to another. This indicates that the reverberation or $RT_{60}$ would vary rapidly in these regions. Since none of these values falls within the JND specified by $P-Reverb$, they cannot be grouped to create regions where reverberation would be imperceptible. This is expected because coupling of spaces is known to affect the sound energy flow and the change of $RT_{60}$ close to the coupling aperture \cite{jing2008visualizations}.

{\small
\begin{table}[t]
        \begin{tabular}{|r|c|c|c|c|c}
        \hline
      {$\mathbf{Region}$} & {$\mathbf{\mu_{mean} (m)}$} &  {$\mathbf{Diff._{max}^{\mu}}$}  & {$\mathbf{RT_{60}^{mean} (s)}$} & {$\mathbf{Diff._{max}^{RT_{60}}}$} \\
        \hline
        
        1 &   2.45  & 1.1 \%  & 0.65 & 4.6 \% \\      
        2 &   3.64  & 0.5 \% & 1.27 & 2.4  \% \\
        3 &   3.11  & 1.1 \%  & 1.03 & 3.6 \% \\
        \hline
        \end{tabular}
        
\caption{\textbf{Mean-Free Path and Reverberation Time Computation:} We show the average values computed using high-order ray tracing for the three different regions shown  in Fig.  \ref{fig:preverbvalidation} and the differences from the average values. Each of these rooms corresponds to imperceptible regions based on our $P-Reverb$ metric. The numerical value shows a maximum variation of $5\%$, which is within JND values of $RT_{60}$. The exact $RT_{60}$ was computed using high-order ray tracing with $300$ bounces. }
\label{table:preverbvalidationtable}
\end{table}
}

\begin{figure}[t]
	\includegraphics[width=0.9\columnwidth]{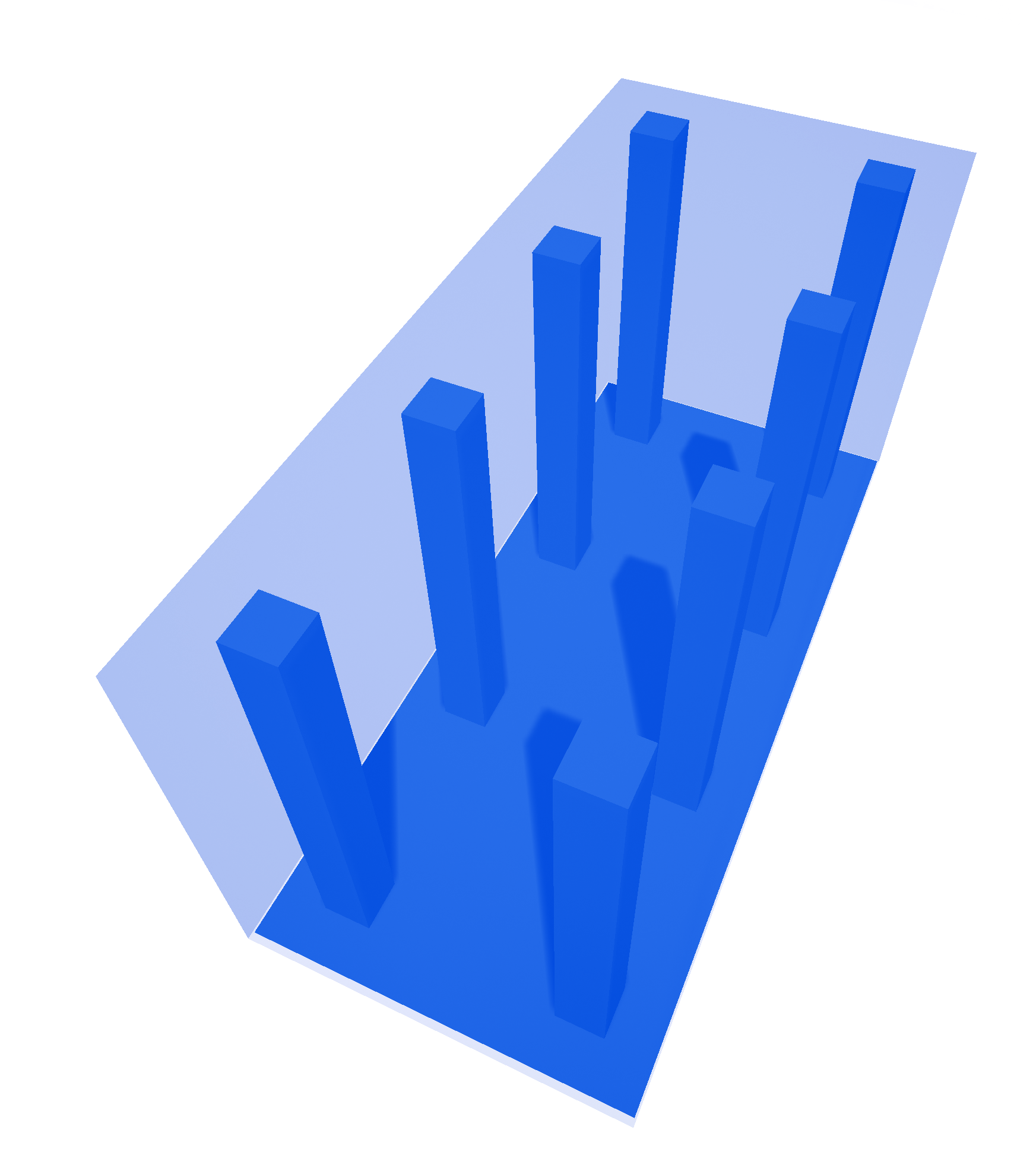}

\caption{Room with Pillars: We illustrate the room with $8$ pillars and use this benchmark to estimate the effectiveness of our mean-free path computation in complex environments with obstacles. We observe less than $3\%$ error using our early reflection based method.}
\label{fig:roomwithpillars}
\vspace*{-0.1in}
\end{figure}

\begin{figure*}[h]
\centering
\includegraphics[width=1\textwidth]{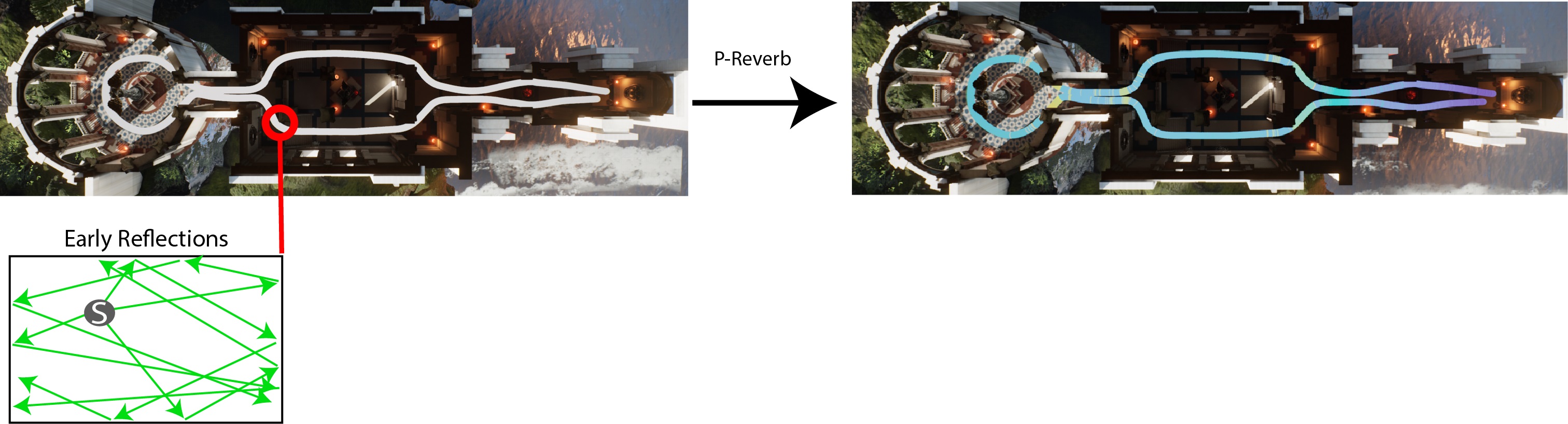}
\caption{The figure shows how the $P-Reverb$ metric can be used to estimate regions where $RT_{60}$ would vary imperceptibly in a scene. The left figure shows a typical listener path in a scene. At each point along this path, we compute the mean-free path $\mu$ using the early reflection based method described. Then using the $P-Reverb$ metric, we cluster the points based on the $JND$ to give us clusters along the path where $RT_{60}$ would vary imperceptibly as shown in the right figure. }
\label{fig:clustering}
\end{figure*}

\section{Interactive Sound Propagation}  
In this section we describe how the $P-Reverb$ metric can be used for interactive sound propagation. As described in Sections 1. \& 2., the sound reaching the listener from a source has three components: direct sound, early reflections, and late reverberation as shown in Fig. \ref{fig:IR_image}. Geometric sound propagation algorithms use methods such as ray tracing to compute the ERs and LRs in the scene. Although early reflections can be computed cheaply, late reverberation computation remains a major bottleneck as it requires very high-order ray bounces in the scene for accuracy making these methods resource heavy. This prevents the use of these methods in interactive environments such as games, which tend to use cheap filter-based approaches (digital reverberation filters) to simulate late reverberation. Reverberation filters require parameters such as $RT_{60}$ to approximate late reverberation in an environment. One way in which reverberation filters can be parameterized accurately is to precompute the $RT_{60}$s along the listener's path in the scene using a high-fidelity geometric sound propagation algorithm such as \cite{schissler2014high}, and then use these precomputed $RT_{60}$ values at runtime in the filter. This would avoid costly high-order ray tracing to simulate reverberation at runtime, but can incur a high precomputation cost requiring us to run high-order ray tracing for every point along the listener's path. We now describe how using our $P-Reverb$ metric can reduce the precomputation cost of computing $RT_{60}$ values in the scene.


\subsection{Sound Propagation using $P-Reverb$} 

\begin{figure}[h]
\centering
\includegraphics[width=1\columnwidth]{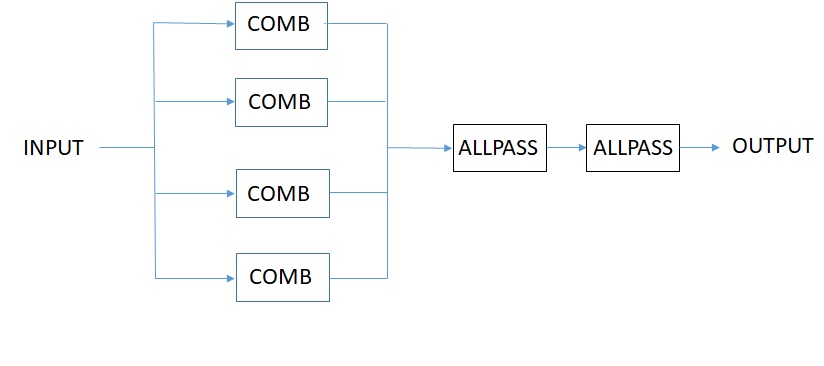}
\vspace*{-0.15in}
\caption{The figure shows the schematic of a typical Schroeder-type filter used in our implementation. The input is processed through a parallel bank of comb filters that create the delayed version of the input signal. The output of this parallel bank goes through a series connection of allpass filters. These filters require parameters like $RT_{60}$ to approximate the late reverberation in a scene.}

\label{fig:filter}
\end{figure}

\subsubsection{Precomputation} We  use our $P-Reverb$ metric to accelerate the pre-computation of late reverberation for an interactive sound propagation system using a Schroeder-type reverb filter to simulate late reverberation (Fig \ref{fig:filter}).  We sample a given scene at multiple points along the listener's path and use a geometric sound propagation method ~\cite{schissler2014high} to compute early reflections by placing an omni-directional sound source tracing $20$ orders of specular reflections at each of these points. Next, using Eq. \ref{eqn:mfp} we compute the mean-free paths at each of these points. Using the $P-Reverb$ metric, we clusters points on the path where $\mu$ varies within its JND, indicating that these regions will have perceptibly similar $RT_{60}$ values (Fig. \ref{fig:clustering}). Finally, using \cite{schissler2014high}, we compute the $RT_{60}$ values once for each computed region using high-order (300 bounces) reflections to get a high quality estimate. Table \ref{table:precomp_tab} shows the speed-up obtained using the $P-Reverb$ metric in precomputation stage. The results were obtained on a multi-core desktop using single thread for the computations.

\subsubsection{Runtime} At runtime, the direct sound computation is done through visibility testing; if a source is visible to the listener, its distance to the listener is used to attenuate the sound pressure according to the inverse distance law. The late reverberation computation is performed using the precomputed $RT_{60}$ values in the previous stage. Given the listener position, a look-up is performed to ascertain the cluster (precomputed in the previous step) the listener position belongs to. Since, an $RT_{60}$ value is associated with each cluster, this is now used as a parameter into the reverberation filter. As long as the listener in within this cluster, $P-Reverb$ metric tells us that $RT_{60}$ value would vary imperceptibly.    
The accompanying video shows the performance of our metric on three different scenes.

\subsection{Benchmarks}

\paragraph{Sun Temple} This scene consists of spatially varying reverberation effects. As the listener moves throughout the scene, the reverberation characteristics vary from being almost dry in the semi-outdoor part of the temple to being reverberant in the inner sanctum. 

\paragraph{Shooter Game} This scene showcases the ability of our method to handle very large scenes. It shows an archetypal video game with multiple levels. As the listener moves from part of the scene to another, it shows our method's ability to handle highly varying, large, virtual environments. 

\paragraph{Tuscany} This scene has two different structures, a house and a cathedral, separated by an outdoor garden. The two structures have very different reverberant characteristics owing to their different geometries, and as the listener moves from the house to the cathedral going through the outdoor garden, the reverberant varies accordingly.  

{\small
\begin{table}[t]
        \centering
        \small
        \begin{tabular}{|r|c|c|c|c|c|c|c}
        \hline
      {$\mathbf{Scene}$} & {$\mathbf{\#Vert.}$} & {$\mathbf{\#P}$} &  {$\mathbf{T_{ER}(ms)}$} &{$\mathbf{T_{LR} (ms)}$} & {$\mathbf{\#P}$}   & {$\mathbf{Speed-up}$} \\
        \hline
        
        Sun Temple      &   215k   &   2301   & 40.2 & 124.2 &  53  & 3x\\
        Tuscany      &   135k   &   1945   &   47.5 & 150.7 & 110   & 3x\\
        Shooter Game      &   49k   &   3235   &  16.7 & 68.4 & 43    & 4x\\
        \hline
        \end{tabular}
        \vspace*{0.1in}
\caption{\textbf{Precomputation Performance Analysis:} We highlight the speed-up in precomputation stage  using the $P-Reverb$ metric. $\#P$ is the number of points along the listener path, $T_{ER}$ is the avg. time taken at each point using ERs,$T_{LR}$ is the average time taken at each point using LRs, and $\#P$ is the number of clusters found using our $P-Reverb$ metric.}
\label{table:precomp_tab}
\end{table}
}

\section{Conclusion, Limitations and Future Work}

 We present a novel perceptual metric that highlights the relationship between the JNDs of early reflections and late reverberation. Our metric is based on two user studies and can be used for fast computation of mean-free-paths and reverberation time in complex environments without high-order ray tracing. Our metric can be used to predict regions in an environment where the reverberation time is likely to vary within its JND value. We evaluate the accuracy of these perceptual metrics and find their accuracy within $5\%$ of the actual values on our benchmarks. 
 
Our approach has some limitations. Our $P-Reverb$ metric computation may not work in totally open environments since the mean-free path computation depends on the presence of collisions with the obstacles in the scene.  Our $P-Reverb$ metric can be regarded as an approximation to a complex function that corresponds to a multi-dimensional perceptual phenomenon dependent on source frequency, scene dimensions, method of sound rendering, material parameters, etc. As a result, we need to perform more evaluations that take other parameters into account. While we observe high accuracy in our current benchmarks, the accuracy could vary in more complex scenes.  Further, our metric tends to be conservative and overestimates the number of regions with similar $RT_{60}$ resulting in running more full simulations than optimal. That being said, it still \changes{significantly reduces the number of full simulations as shown in Table \ref{table:precomp_tab}}. \changes{Our experimental work also has limitations, including the restricted range of room sizes (motivated by the psychophysical goal), the fixed listener, and the restriction to mono rendering.} As part of future work, we would like to overcome these limitations and further evaluate our approach on complex scenes and use them for multi-modal rendering.

\section{Acknowledgments} The authors would like to thank the subjects who took part in
the user-study. This work was supported in part by ARO grant W911NF-18-1-0313, NSF grant 1840864, and Intel.

\bibliographystyle{abbrv-doi}
\bibliography{main}
\end{document}